\documentclass[letterpaper,twocolumn,showpacs,preprintnumbers,amsmath,amssymb,floatfix]{revtex4}

\usepackage{graphicx}
\usepackage{dcolumn}
\usepackage{bm}
\usepackage{epsfig}

\begin{document}

\title{Synchronization from Disordered Driving Forces in Arrays of Coupled Oscillators}

\author{Sebastian F.~Brandt}
\email{sbrandt@physics.wustl.edu}
\author{Babette K.~Dellen}
\author{Ralf Wessel}
\affiliation{
Department of Physics, Campus Box 1105, Washington University in St.~Louis, MO 63130-4899, USA}
\date{December 21, 2005}
\begin{abstract}
The effects of disorder in external forces on the dynamical behavior of coupled nonlinear oscillator networks are studied. When driven synchronously, i.e., all driving forces have the same phase, the networks display chaotic dynamics. We show that random phases in the driving forces result in regular, periodic network behavior. Intermediate phase disorder can produce network synchrony. Specifically, there is an optimal amount of phase disorder, which can induce the highest level of synchrony. These results demonstrate that the spatiotemporal structure of external influences can control chaos and lead to synchronization in nonlinear systems.
\end{abstract}
\maketitle
Networks of coupled nonlinear oscillators provide useful model systems for the study of a variety of phenomena in physics and biology \cite{Heagy}. Among many others, examples from physics include solid state lasers \cite{Roy} and coupled Josephson junctions \cite{Ustinov,Wiesenfeld1}.  In biology, the central nervous system can be described as a complex network of oscillators \cite{Amit}, and cultured networks of heart cells are examples of biological structures with strong nearest-neighbor coupling \cite{Soen}.  In particular, the emergence of synchrony in such networks \cite{Pikovsky,Strogatz} and the control of chaos in nonlinear systems \cite{OGY,Pyragas1,Pyragas2} have received increased attention in recent years.

Disorder and noise in physical systems usually tend to destroy spatial and temporal regularity. However, in nonlinear systems, often the opposite effect is found and intrinsically disordered processes, such as thermal fluctuations or mechanically randomized scattering, lead to surprisingly ordered patterns \cite{Shinbrot}.  For instance, in the phenomenon of stochastic resonance the presence of noise can improve the ability of a system to transfer information reliably \cite{Wiesenfeld2}.
Some time ago, Braiman et al.~studied one- (1D) and two-dimensional (2D) coupled arrays of forced, damped, nonlinear pendula \cite{Braiman}. They found that when a certain amount of disorder was introduced by randomizing the lengths of the pendula the dynamics of the array ceased to be chaotic.  Instead, they observed complex, yet regular, spatiotemporal patterns.  Further studies of the same system showed that chaos in the array of oscillators can also be tamed by impurities \cite{Geisel} and that random shortcuts between the pendula lead to synchronization of the array \cite{Feng}.
 
Here, we introduce disorder by modifying the driving forces of the oscillators through phase differences.  We observe the emergence of regular, phase-locked dynamics.  Moreover, for intermediate spreads of the phase angles in the driving forces, we find that the oscillations become largely synchronous.

We focus our numerical analysis on arrays of forced, damped, nonlinear pendula. The 1D array (chain) is described by the equation of motion
\begin{eqnarray}
ml^2\ddot{\theta}_n + \gamma \dot{\theta}_n &=& - mgl \sin \theta_n
+ \tau' + \tau \sin \left( \omega t + \varphi_n \right) \nonumber \\ && \hspace{-25mm} +
\kappa(\theta_{n+1} + \theta_{n - 1} - 2 \theta_n), \quad \quad n = 1, 2, \ldots N \, .  \label{eq1}
\end{eqnarray} 
In order to consider a 2D lattice, we introduce an additional index, $\theta_n \rightarrow \theta_{n,m}, \, n, \, m = 1, 2, \ldots N$ and modify the coupling term accordingly: $\kappa(\theta_{n+1} + \theta_{n - 1} - 2 \theta_n) \rightarrow \kappa(\theta_{n+1,m} + \theta_{n - 1,m}+ \theta_{n,m+1} + \theta_{n,m-1} - 4 \theta_{n,m})$.  For both the 1D and 2D case, we choose free boundary conditions, i.e., $\theta_0 = \theta_1, \, \theta_N = \theta_{N+1}$ and $\theta_{0,m} = \theta_{1,m}, \, \theta_{N,m} = \theta_{N+1,m}, \, \theta_{n,0}=\theta_{n,1},\, \theta_{n,N} = \theta_{n,N+1}$, respectively. 
The parameter values used are the same as in previous studies \cite{Braiman,Geisel,Feng}: The mass of the  pendulum bob is $m = 1$, the length $l = 1$, the acceleration due to gravity $g = 1$, the damping $\gamma = 0.75$, the d.c.~torque $\tau' = 0.7155$, the a.c.~torque $\tau = 0.4$, the angular frequency $\omega = 0.25$, and the coupling strength $\kappa = 0.5$. For this choice of parameter values, each isolated pendulum displays chaotic behavior characterized by a positive Lyapunov exponent \cite{Braiman}. 

A particularly easy and intuitive way to visualize the global spatiotemporal behavior of a chain (or lattice) of oscillators is to consider the average velocity
\begin{eqnarray}
\sigma(jT) = \frac{1}{N} \sum_{n = 1}^N \dot{\theta}_n(jT) \label{eq2} 
\end{eqnarray}
at times that are integer multiples of the forcing period $T = 2 \pi / \omega$ \cite{Geisel}. Considering this measure for an isolated pendulum, Gavrielides et al.~perfomed a bifurcation analysis with respect to the pendulum length $l$ and found that an uncoupled pendulum is chaotic for values $l = 1 \pm 0.002$ \cite{Gavrielides}.
If the length of an isolated pendulum is increased to $l > 1.002$, it performs a `libration,' in which the combined d.c.~and a.c.~torque are insufficient to overcome the pendulum's increased rotational inertia.  On the other hand, if the pendulum's length is decreased to $l < 0.998$, the pendulum performs a `rotation,' an overturning motion where the torques combine to rotate the pendulum over the top.
  
\begin{figure}[t]
\centerline{\includegraphics[width=\columnwidth]{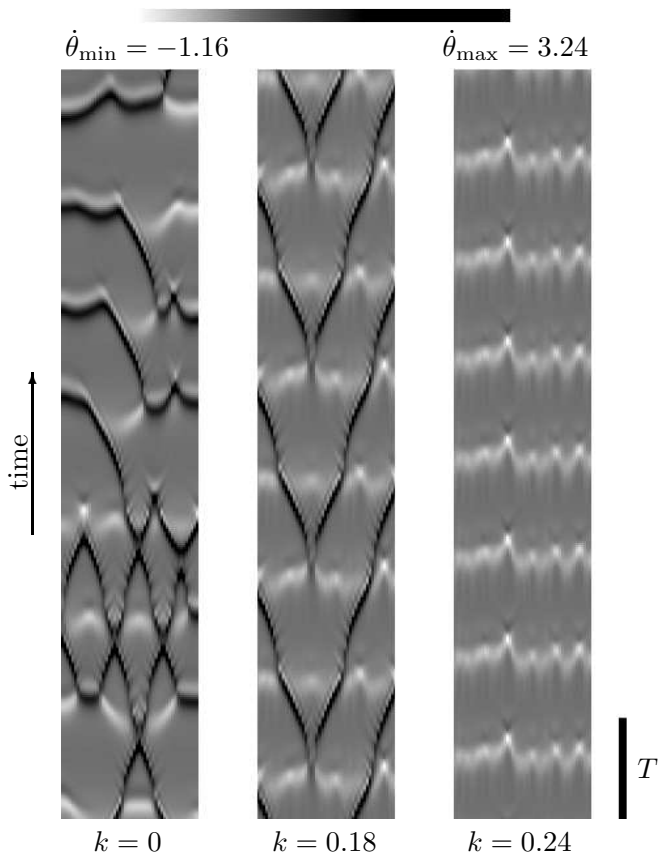}}
\caption{Spatiotemporal angular velocity plots for chaotic and regular dynamics in an array of $N = 50$ coupled oscillators. The chain of pendula is shown from left to right. Time increases continuously from bottom to top. Grayscales indicate the angular velocities of the oscillators. Light gray shades represent negative, dark tones positive velocities.}
\label{fig1}
\end{figure}

In our study, we do not alter any parameters that would affect the dynamics of an isolated pendulum and keep the coupling strength at its default value.  
Instead, we introduce disorder by randomly varying the phase angles $\varphi_n$ of the driving forces in Eq.~(\ref{eq1}). 
In the case where $\varphi_n = 0$ for all driving forces, we observe chaotic dynamics in the array (Fig.~\ref{fig1}) in agreement with previous studies \cite{Braiman}.
However, when we disorder the driving forces by randomly choosing the phase angles $\varphi_n$ 
uniformly from the interval $[-k \pi, + k \pi]$, we observe that 
for sufficiently large $k$ the oscillations become regular. 

\begin{figure}[t]
\centerline{\includegraphics[width=\columnwidth]{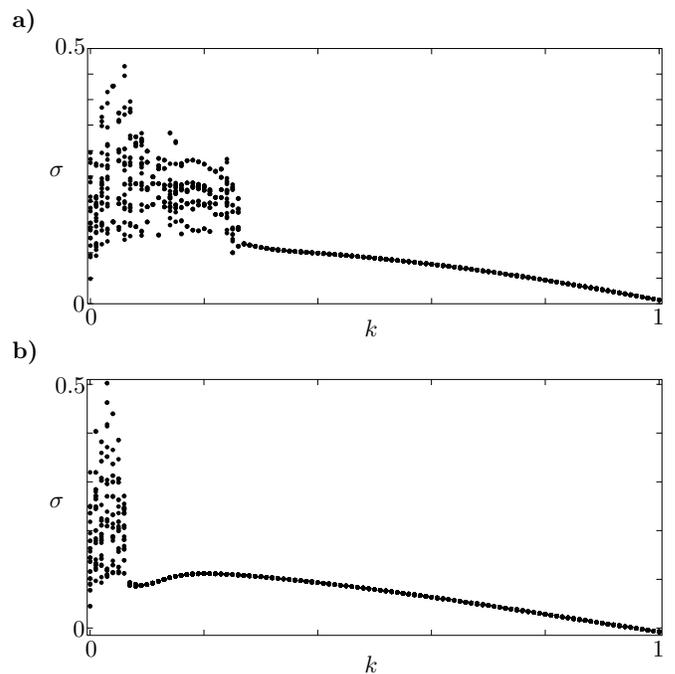}}
\caption{Chaotic and regular dynamics as a function of the degree of disorder. The average angular velocity at $t = 60T,\, 61T, \, \ldots, \, 80 T$ is shown for each value of the disorder parameter $k$. {\bf a)} 1D array of $N=50$ oscillators. {\bf b)} 2D lattice of $16 \times 16$ oscillators.}
\label{fig2}
\end{figure}

Figure \ref{fig2} shows the average angular velocity $\sigma(t)$ at $t = 60 T,\, 61 T, \, \ldots,\, 80 T$ for a 1D array of $N = 50$ and a 2D lattice of $16 \times 16$ oscillators.
The presence of chaos for small disorder in both the 1D and 2D array becomes manifest in a dispersed distribution of the average velocities $\sigma(60T), \, \sigma(61T), \, \ldots, \,  \sigma(80T)$. For larger disorder, however, we observe periodic patterns in the form of $1T$-,\linebreak[4] $2T$-, $3T$-, $\ldots$ `attractors,' where the average velocity of the oscillator array repeats its value after 1, 2, 3, $\ldots$ forcing periods. Ultimately, as $k$ is increased further, a $1T$ periodic pattern is reached.

\begin{figure*}[t]
\begin{minipage}{\columnwidth}
\centerline{\includegraphics[width=\columnwidth]{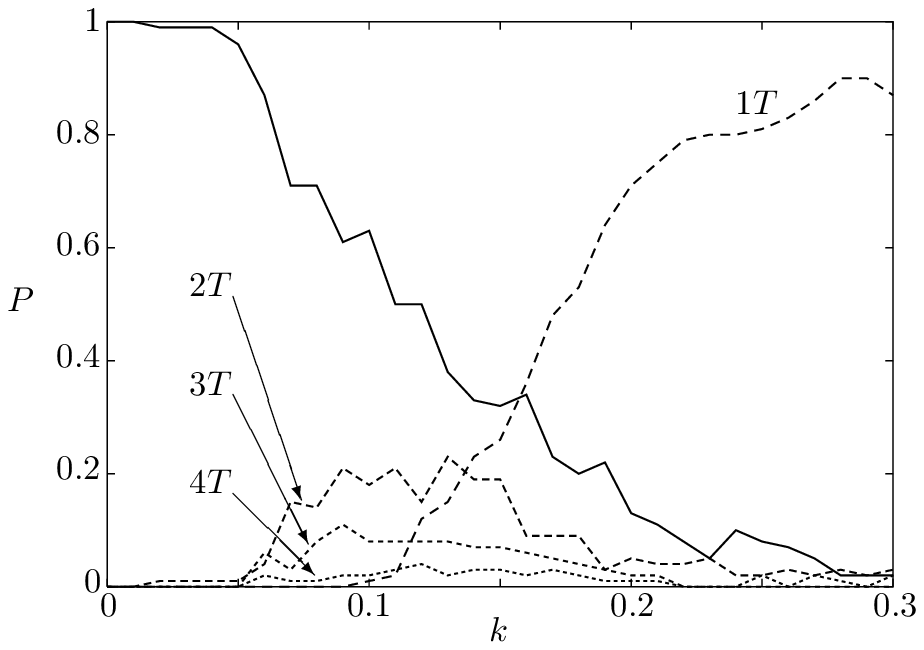}}
\caption{Probability $P$ of chaotic dynamics (solid line) and different forms of regular behavior (dashed lines) vs.~the disorder parameter $k$ in an array of $N = 50$ coupled oscillators.  The probabilities were determined by averaging over $100$ different samplings of the phases $\varphi_n$.}
\label{fig3}
\end{minipage}
\hfill
\begin{minipage}{\columnwidth}
\centerline{\includegraphics[width=\columnwidth]{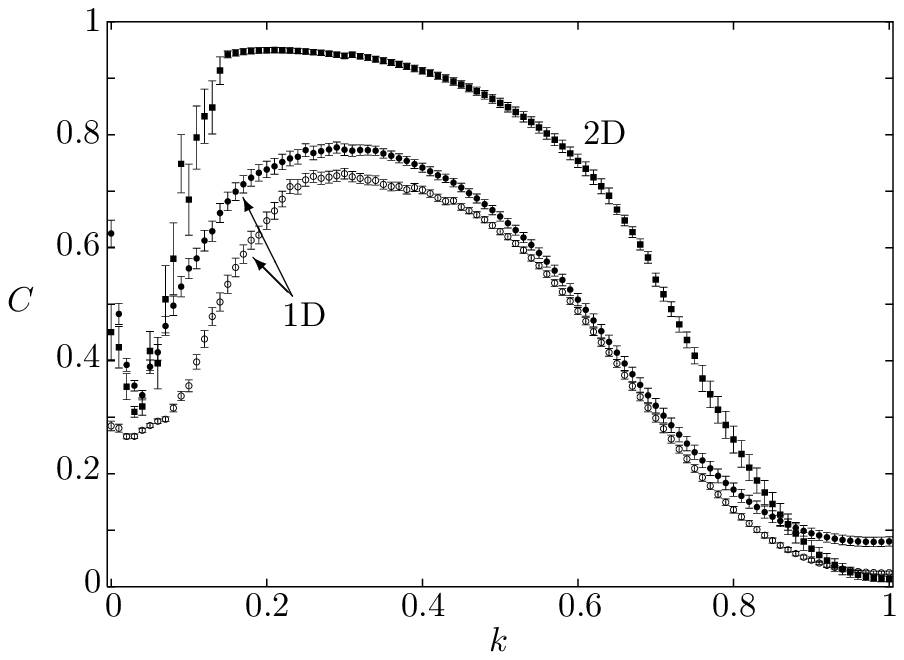}}
\caption{Synchronization in 1D (dots) and 2D (squares) arrays of oscillators vs.~the disorder parameter $k$. Error bars show one standard error of the mean. Filled dots correspond to $N=16$, open dots to $N=50$. Averaging was performed over 200 (1D) and 10 (2D) different samplings of the phases $\varphi_n$. }
\label{fig4}
\end{minipage}
\hfill
\end{figure*}

In general, the value of $k$ for which a transition from chaotic to regular dynamics first occurs depends on the particular distribution of the random phases.  We thus consider the average over several different samplings of uniform distributions in order to analyze the occurrence of different forms of periodic behavior.
Figure \ref{fig3} shows the probability for a 1D array to have reached a $1T$-, $2T$-, $3T$-, or $4T$-attractor after $t=60T$ as a function of the disorder parameter $k$.
For very small disorder, i.e., $k < 0.02$, we observe only chaotic dynamics, but as $k$ passes this threshold, the first periodic patterns start to appear.  For $k \geq 0.1$, we observe that $1T$-, $2T$-, $3T$-,\linebreak[4] $4T$-, $\ldots$ attractors coexist with chaotic behavior. For $0.02 \leq k \leq 0.13$ the $2T$-attractor is the dominant form of dynamics if an attractor has been reached.
For $k > 0.28$, the array undergoes regular oscillations with period $1T$ in the vast majority of cases.

Furthermore, in addition
to the transition from chaotic to regular behavior, we observe that the oscillations become largely synchronous, i.e., the phases of the oscillations not only lock but tend to assume equal values, for intermediate values of $k$.  In order to quantify the presence of synchrony in the array, we consider the averaged cross correlation
\begin{eqnarray}
C = \frac{2}{N(N-1)}\sum_{i<j}c_{ij}\,,
\end{eqnarray}
where $c_{ij}$ denotes the correlation between the $i$th and $j$th oscillator:
\begin{eqnarray}
c_{ij} = \frac{\int_{T_0}^{T_0 + T} dt \, \dot{\theta}_i(t) \dot{\theta}_j(t)}{\left[\int_{T_0}^{T_0 + T} dt \, \dot{\theta}_i^2(t) \int_{T_0}^{T_0 + T} dt \, \dot{\theta}_j^2(t)\right]^{1/2}} \,. 
\end{eqnarray}
Figure \ref{fig4} shows $C$ as a function of $k$ for two 1D and one 2D arrays. 
Disordering the driving forces results in less synchronized oscillations of the array if the disorder parameter is very small. The minimum of synchrony is reached for $k\approx 0.03$.  Note that the location of this  minimum corresponds approximately to the first appearance of regular dynamics in Fig.~\ref{fig3}. 
When the external forces are disordered further, synchronization in the array increases and reaches a peak value for intermediate disorder.  In the 1D case, the maximum is reached for $k \approx 0.3$ and its value is $C_{\rm max} \approx 0.72$ for $N=50$ and $C_{\rm max} \approx 0.78$ for $N=16$ oscillators.  In the case of the 2D array, the synchronization is even stronger.  
Here, the peak value of $C_{\rm max} \approx 0.95$ is reached for $k \approx 0.2$. We attribute the stronger synchronization in the 2D array to the fact that the number of couplings per oscillator is higher than in the 1D case. Furthermore, smaller arrays show a higher degree of averaged cross correlation than larger arrays.  This is because oscillators that are nearest neighbors show the highest degree of synchronization, and the ratio of cross-correlation coefficients obtained from direct neighbors to all cross-correlation coefficients contributing to the averaged cross correlation $C$ decreases with increasing size of the array like ${\cal O}(1/N)$.

To summarize, we have shown that disorder leads to transitions from chaotic to regular behavior in arrays of coupled oscillators when disorder is introduced in the phases of the driving forces \cite{fn1}. In this investigation, each pendulum was in a regime where it behaves chaotically when uncoupled, in contrast to previous studies in which parameters were altered that affect the dynamics of an isolated oscillator \cite{Braiman,Geisel}.  
In particular, Braiman et al.~introduced disorder by randomly varying the lengths of the pendula \cite{Braiman}. 
Since an isolated pendulum only behaves chaotically when its length lies within a narrow range, only 2\% of the oscillators remained in their chaotic regime in this approach, and the transition from chaotic to regular spatiotemporal patterns reported in Ref.~\cite{Braiman} can be attributed to the dominance of the majority of regular pendula over the few remaining chaotic ones \cite{Geisel}. 
Our results show that disorder in the model system described by Eq.~(\ref{eq1}) results in regular dynamics of the array even if all individual elements are chaotic. Moreover, we find that for intermediate disorder, the oscillations show a high degree of synchronization.  

Stimulus-induced synchronization of neural activity in central nervous systems has intrigued neuroscientists for decades \cite{Ritz,Buzsaki}. Furthermore, in many applications, such as in coupled Josephson junctions, or in the case of atrial or ventricular fibrillation, one seeks to restore periodic or steady-state behavior from chaos. It is in regard to these day-to-day circumstances that control and synchronization of chaotic dynamics have become one of the central topics of nonlinear science \cite{Boccaletti,ChaosFocIss}.
In most situations the components of a system themselves cannot be altered, so it is desirable to establish methods by which chaos can be tamed without changing parameters intrinsic to the system.  We thus believe that our proposed mechanism of controlling chaos via external forces has potential applications in these fields.

We thank Kevin Archie, Anders Carlsson, and John Clark for critical reading of the manuscript. This work was supported by NIH-EY 15678.


%

\begin{thebibliography}{99}
\bibitem{fn1} The observed transition also occurs in other systems where the potential that governs the dynamics of a single oscillator is periodic.  If, for instance, we choose a potential of the form $V(\theta) = - \alpha \cos^4(\theta/2)$ instead of the choice $V(\theta) = - m g l \cos(\theta)$ in Eq.~(1), we obtain similar results for a suitable choice of the parameter $\alpha$ (data not shown).
\end{thebibliography}
\end{document}